\documentclass[12pt]{article}
\usepackage{graphicx}

\def\beq {\begin{equation}}
\def\eeq {\end{equation}}
\def\bea {\begin{eqnarray}}
\def\eea {\end{eqnarray}}
\def\ni {\noindent}

\def\rar {\rightarrow}

\def\sb {\bar{s}}

\def\Kb {\bar{K}}

\def\p {\pi}

\def\r {\rho}

\def\dkpp0{D^+\to \bar{K^*_0}\p^0\p^+}

\def\dkpp{D^+  \to K^-\pi^+\pi^+ }

\def\pbnr{}
\def\speaker{\underline{Patr{\'i}cia Camargo Magalh{\~a}es} and Manoel Roberto Robilotta}
\def\onbehalfof{ }
\def\title{A complete model for $\dkpp$ s-wave }
\def\affiliation{Institute of Physics, University of S{\~a}o Paulo, Brazil\\
}
\def\support{This work was supported by FAPESP, Brazilian founding agency.}

\newcommand\pubnumber{\pbnr}
\newcommand\pubdate{\today}
\def\Title#1{\begin{center} {\Large #1 } \end{center}}
\def\Author#1{\begin{center}{ \sc #1} \end{center}}

\newcommand{\OnBehalf}[1]{\sbox0{#1}\ifdim\wd0=0pt
        {}
	\else
	{\\on behalf of #1}
	\fi}
\newcommand{\SupportedBy}[1]{\sbox0{#1}\ifdim\wd0=0pt
        {}
	\else
	{\footnote{#1}}
	\fi}
\def\Address#1{\begin{center}{ \it #1} \end{center}}

\newcommand\pubblock{\includegraphics[width=5cm]{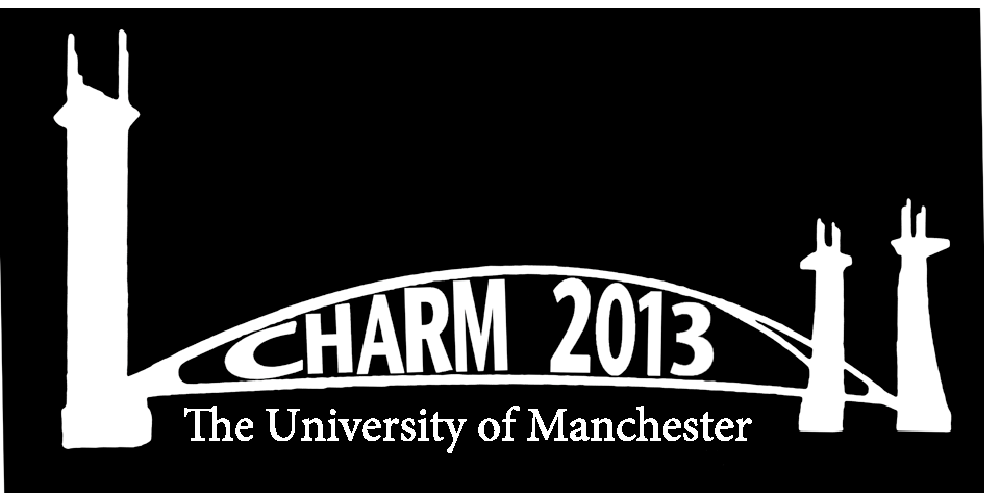}\hfill{\begin{tabular}{l} \pubnumber\\
         \pubdate  \end{tabular}}}
\newenvironment{Abstract}{\begin{quotation}  }{\end{quotation}}
\newenvironment{Presented}{\begin{quotation} \begin{center} 
             PRESENTED AT\end{center}\bigskip 
      \begin{center}\begin{large}}{\end{large}\end{center} \end{quotation}}
\def\Acknowledgements{\bigskip  \bigskip \begin{center} \begin{large}
             \bf ACKNOWLEDGEMENTS \end{large}\end{center}}
\def\venue{The 6$^{th}$ International Workshop on Charm Physics\\
(CHARM 2013)\\
Manchester, UK,  31 August -- 4 September, 2013}

\def\beq{\begin{equation}}
\def\eeq#1{\label{#1}\end{equation}}
\def\eeqn{\end{equation}}

\def\beqa{\begin{eqnarray}}
\def\eeqa#1{\label{#1}\end{eqnarray}}
\def\eeqan{\end{eqnarray}}

\let\bar=\overbar

\def\Dslash{\not{\hbox{\kern-4pt $D$}}}
\def\dslash{\not{\hbox{\kern-2pt $\del$}}}

\def\msb{{\bar{\ssstyle M \kern -1pt S}}}

\begin{document}
\begin{titlepage}
\pubblock

\vfill
\Title{\title}
\vfill
\Author{\speaker\SupportedBy{\support}\OnBehalf{\onbehalfof}}
\Address{\affiliation}
\vfill
\begin{Abstract}
The $\dkpp$ decay have been observed in LHCb collaboration with 
thousands of events per second from $K\p$ threshold. The theoretical treatment of this decay includes a rich dynamic behaviour that mix weak and strong interactions in a non trivial way. 

In the present work we show recently progress on $\dkpp$ decay within a effective hadronic formalism. The weak sector followed a similar formalism from heavy meson ChPT\cite{HM}, where quark c  is an external object of  SU(3) Goldstone bosons sector and  FSI description is based in three-body model from\cite{BR}.
The results concerning  the vector weak transition revealed  a very interesting new scenario in  s-wave phase-shift: agrees well with data in the elastic region, and it starts from $-60$ degrees. Up to now the reason way the experimental phase starts at negative degrees($\approx - 138$) was a mystery. However, we show that this behaviour is a consequence of interference between a correct weak vertex dynamic, dominated by $\rho$ meson propagator, and a rescattering  model to the final state interaction.
Therefore, this work represent a progress in the understanding of $ \dkpp$ and could be applied to other decays.
\end{Abstract}
\vfill
\begin{Presented}
\venue
\end{Presented}
\vfill
\end{titlepage}
\def\thefootnote{\fnsymbol{footnote}}
\setcounter{footnote}{0}

\section{Introduction}
The $S$-wave $K^- \p^+$ sub-amplitude in the decay 
$D^+ \rar K^- \p^+ \p^+$, denoted by $[K^-\p^+]_{D^+}$
has been extracted from data by the E791\cite{E791k} 
and FOCUS\cite{FOCUS} collaborations.
A remarkable feature of the results is a  significant phase shift deviation 
between $[K^-\p^+]_{D^+}$ and elastic $K^- \p^+$ LASS data \cite{LASS}, which
 was considered to be a puzzle until recently. However, this is a clearly indication that the dynamical relationship between
both types of processes is not simple and has motivated 
an effort by our group aimed at understanding the
origins of this problem. A schematic calculation was presented in \cite{BR} and recent 
progress can be found in \cite{Buz}. Here we briefly summarize the main issues and progress.
The programme is implemented by means of effective lagrangians,
which incorporate the symmetries of QCD, where weak and electromagnetic 
interactions are included as external sources.
The inclusion of heavy mesons can be performed by suitable adaptations
of the light sector \cite{HM}.
%
\section{Dynamics}
\begin{figure}[htb]
\centering
\includegraphics[width=0.6\columnwidth,angle=0]{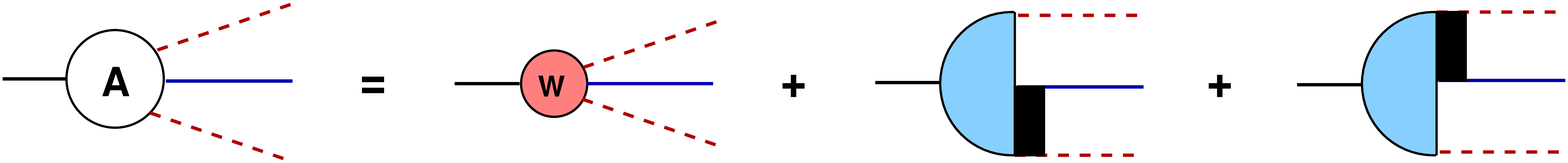}
\\[3mm]
\includegraphics[width=.75\columnwidth,angle=0]{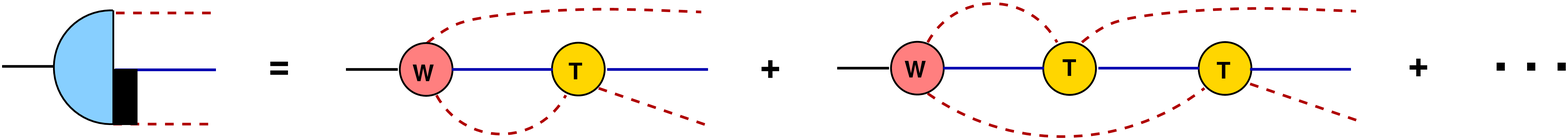}
\caption{Diagrammatic representation of the heavy meson decay
 into $K\pi\pi$, starting from the weak amplitude (red)
and including hadronic final state interactions.}
\label{FAmp}
\end{figure}

The reaction $D^+ \rar K^- \p^+ \p^+$ involves two distinct structures: weak vertex and final state interactions (FSI).
The first one concerns the primary quark transition $c\rar s \;W^+\,$,
which occurs in the presence of the light quark condensate 
of the QCD vacuum and is dressed into hadrons.
The second class of processes corresponds to three-body
FSI, 
associated with the strong propagation of the state produced 
in the weak vertex to the detector.
These ideas are summarized in Fig.\ref{FAmp}.
\subsection*{weak vertex}
\begin{figure}[htb] 
\begin{center}
\vspace{-1mm}
\includegraphics[width=.6\columnwidth,angle=0]{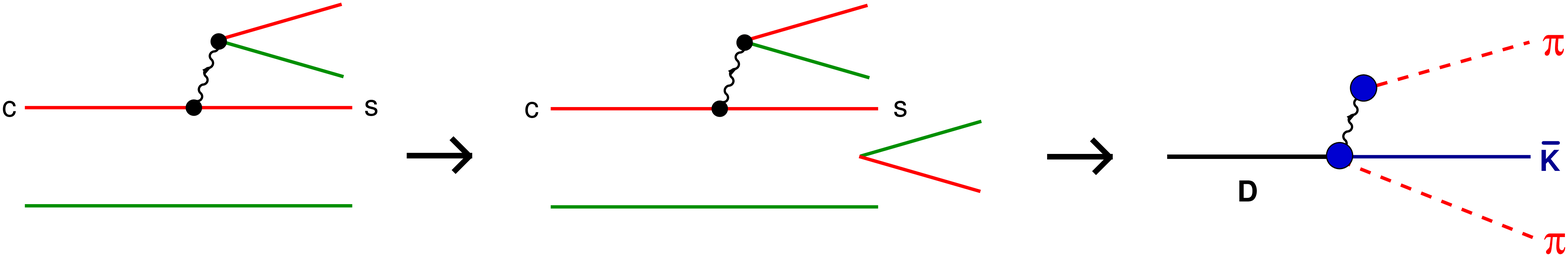}\\[4mm]
\includegraphics[width=.6\columnwidth,angle=0]{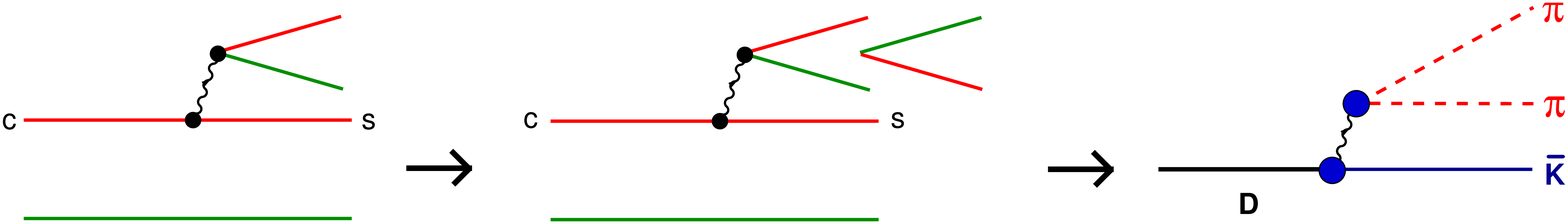}
\caption{Mechanisms for hadronization; quarks $u$ and $d$ are 
not specified.}
\end{center}
\label{FChaoH}
\end{figure}

For description of the weak vertex 
in $D^+ \rar K^- \p^+ \p^+$,
 we concentrate  on the colour allowed class of diagrams proposed by Chau\cite{Chau}, which 
gives rise to the hadronic amplitudes shown in Fig.\ref{FChaoH}.
Processes on the top involve
an {\em axial} weak current, whereas the bottom diagram is based on 
a {\em vector} current. 
The blob in the diagrams summarizes several hadronic 
processes which contribute to form factors.
%
%
In the absence of form factors, the  weak vertex
entering Fig.~\ref{FAmp} is given by the diagrams shown in Fig. 3, 
where processes $a$ and $b$ involve the axial current and $c$
contains a vector  current.
%
\begin{figure}[htb]
\begin{center}
\includegraphics[width=.7\columnwidth,angle=0]{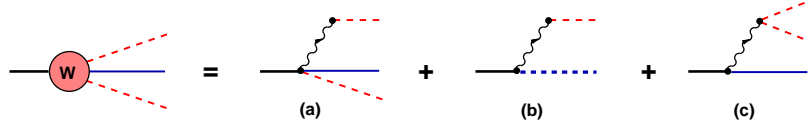}
\caption{Topologies for the weak vertex: the dotted line is a scalar 
resonance and the wavy line is a $W^+$, which is contracted to a point 
in calculations.}
\end{center}
\label{fig:FAmpW}
\end{figure}
\ni The inclusion of form factors can be made either 
by using phenomenological input or by  allowing the intermediate 
propagation of $(c\sb)$ states, as shown in Fig. 4. 
%
\begin{figure}[htb] 
\begin{center}
\vspace*{-2mm}
\includegraphics[width=.7\columnwidth,angle=0]{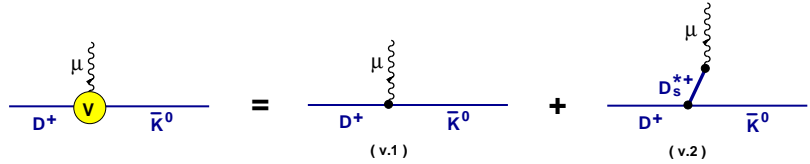}
\caption{Vector form factor.}
\end{center}
\label{FVFF}
\end{figure}
\subsection*{final state interactions}
When final state interactions are added  to the processes, 
one finds three families of color-allowed amplitudes denoted respectively by $A_a\,$, $A_b\,$, and $A_c\,$.
The class of FSIs considered is based on a succession of elastic 
two-body interactions, which bring the $K \p$ phase into the problem.
The $K \p$ amplitude 
is derived by means of chiral effective lagrangians,
based on leading order contact terms \cite{GL} and 
supplemented by resonances \cite{EGPR}, which allow for a wider 
energy range.

\section{First results}
In a previous publication \cite{BR}, we evaluated the contributions
of the three topologies (Fig. 3
) up to second order in a perturbative series
 to the $S$-wave $[K^-\p^+]_{D^+}$.
With the purpose of taming the calculation we made some simplifying assumptions:
the weak amplitudes of Fig. 3  were taken to be constants,
isospin $3/2$ and $P$ waves were not included in the $K\p$ amplitude
and couplings to either vector mesons or inelastic channels
were neglected.
%
\begin{figure}[htb]
\vspace*{-3mm}
\begin{center}
\includegraphics[width=0.6\columnwidth,angle=0]{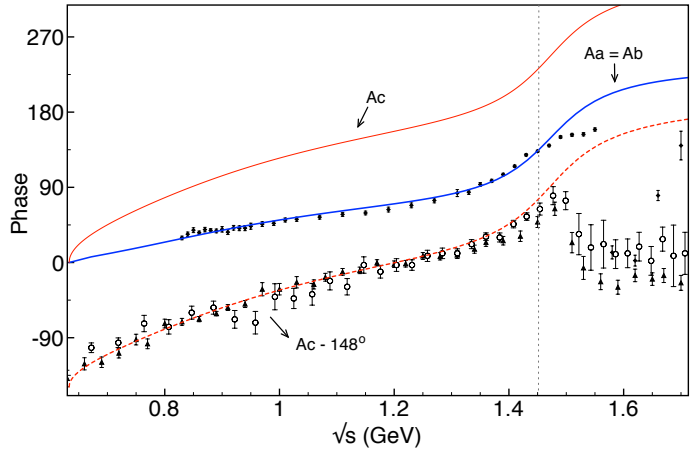}
\caption{Phase for $A_a$, $A_b$, $A_c$ at leading order and $A_c$ shifted by $-148^0$ compared with FOCUS 
\cite{FOCUS}(triangle) and E791\cite{E791k}(circle) data, together with elastic $K\pi$ results from LASS\cite{LASS}(diamond).}
\end{center}
\label{FShifted}
\end{figure}
%
Results for the phase are displayed in Fig. 5. 
Leading order contributions from the axial weak current,
represented by $A_a$ and $A_b\,$, obey Watson's theorem 
and fall on top of elastic $K\p$ data.
The curve for the vector $A_c$ amplitude
has a different shape and, if shifted by $-148^0$,
can describe well FOCUS data\cite{FOCUS}, up to the region of the peak.

The main lesson to be drawn from our first approach to this 
problem is that, for some yet unknown reason,
the amplitude which begins with a vector weak current, 
represented by diagram $(c)$ of Fig. 3, 
seems to be favoured by data.
This amplitude receives no contribution at tree level, since the
$W^+$ emitted by the charmed quark decays into a $\p^+\p^0$ pair.
Therefore, the leading term in this kind of process necessarily
involves loops and the corresponding imaginary components.

\section{Vector vertex}

A limitation of our first study \cite{BR} 
 was that all weak vertices 
were described by momentum independent functions.
Those results are now improved by considering the proper $P$-wave 
structure of the weak vertex, 
corrections associated with form factors and   
contributions from intermediate $\rho$ mesons.
\begin{figure}[htb] 
\begin{center}
\includegraphics[width=0.3\columnwidth,angle=0]{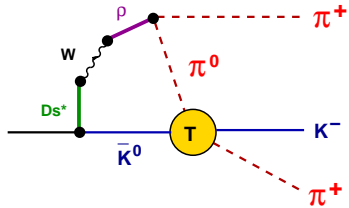}
\caption{Leading vector contribution.}
\end{center}
\label{Fvec}
\end{figure}
%
The $\rho$ is introduced by means of standard vector meson dominance,
using the formalism given in Ref.\cite{EGPR}. 
The $D \rar W \Kb$ vertex may contain $(\sb c)$ intermediate states 
and can be obtained either by means of heavy-meson effective 
lagrangians \cite{HM} and the diagrams of 
Fig.~4, 
or by using phenomenological information
parametrized in terms of nearest pole dominance \cite{weakFF}. Our basic interaction for this process became the diagram in Fig.~6.
%
\begin{figure}[htb]
\begin{center}
\hspace*{-25mm}
\includegraphics[width=0.55\columnwidth,angle=0]{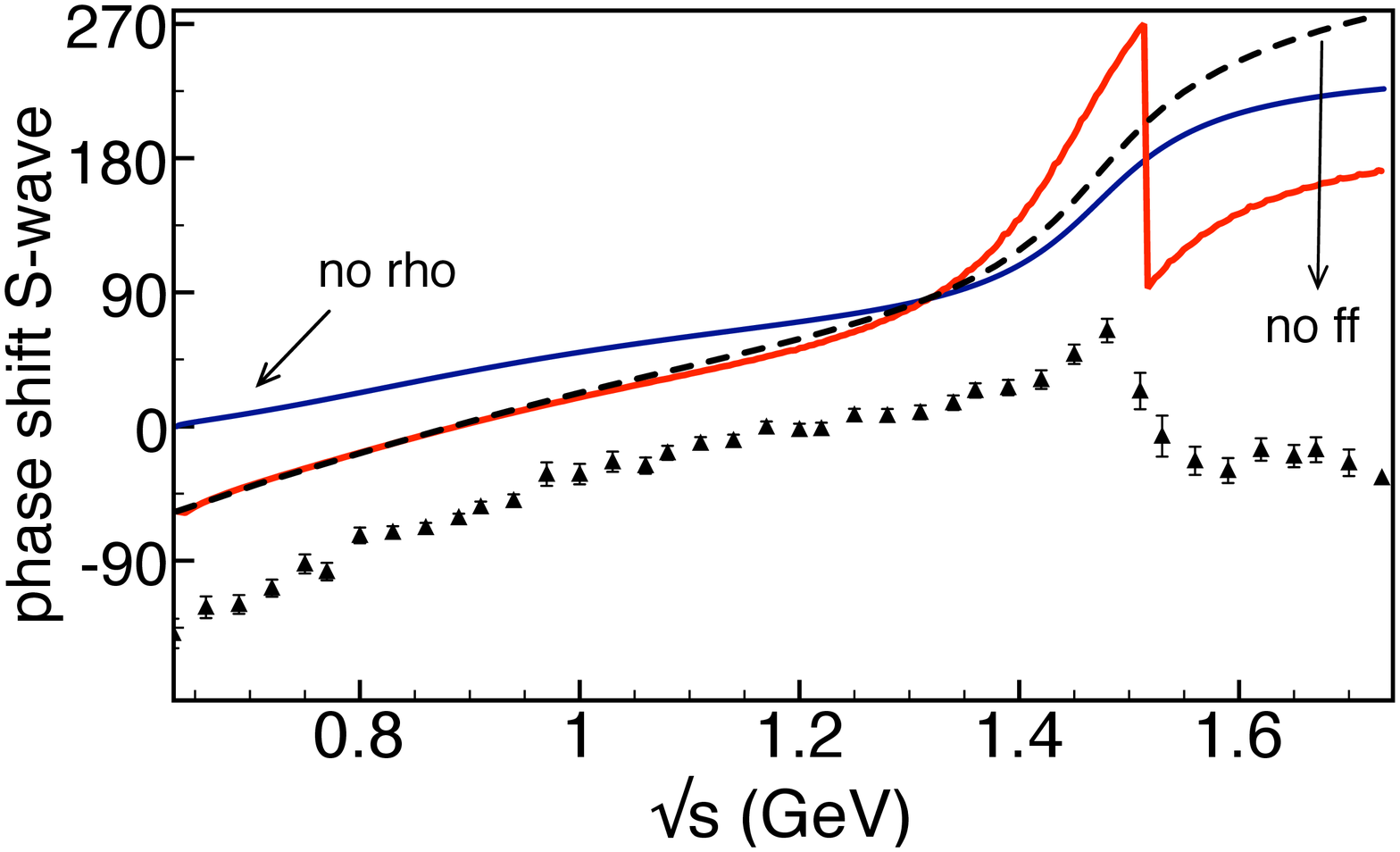}
\hspace*{2mm}
\includegraphics[width=0.55\columnwidth,angle=0]{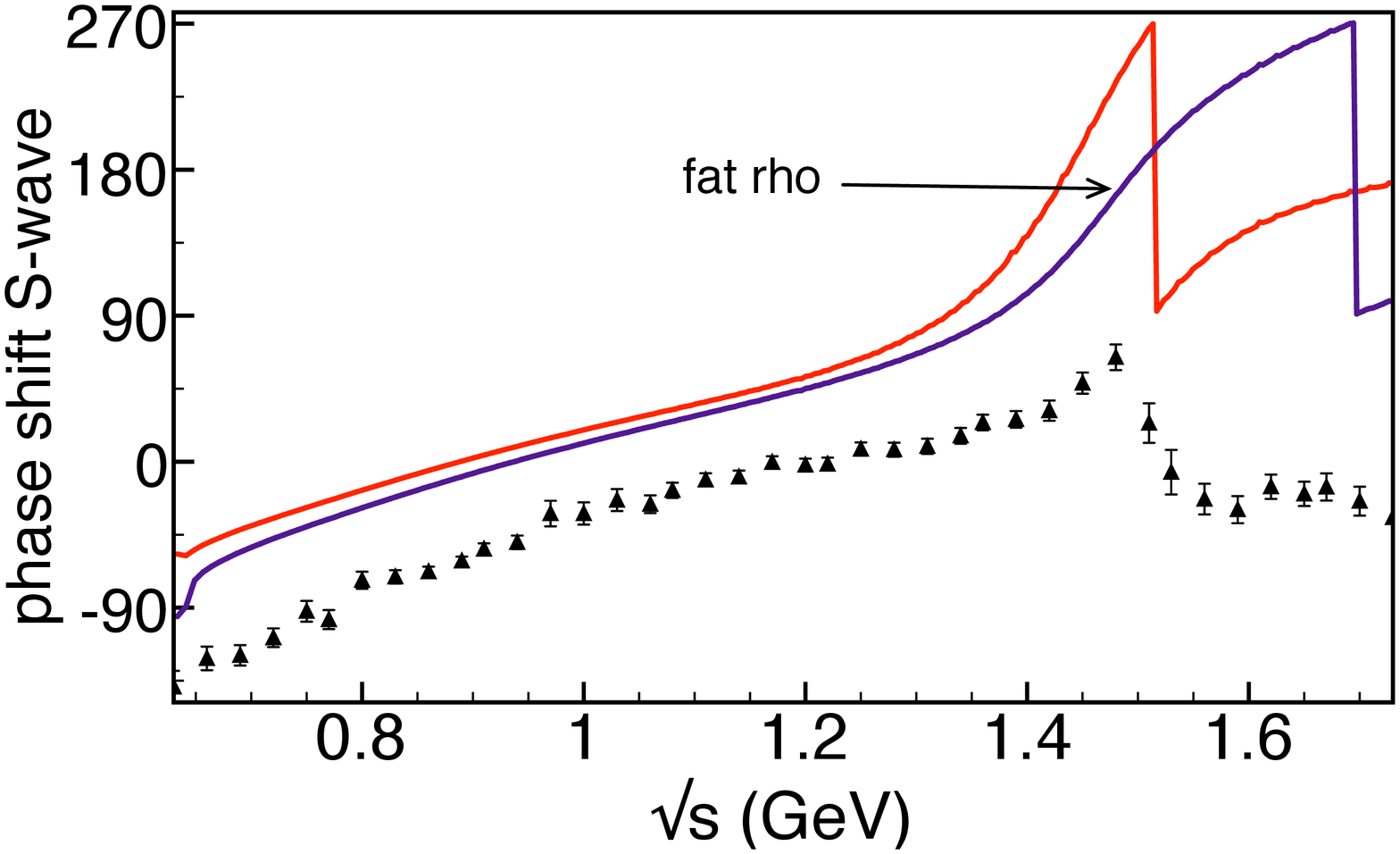}
\caption{Predictions for the phase, compared with FOCUS 
results\cite{FOCUS}.}
\end{center}
\label{FNPh}
\end{figure}
%
New predictions for the phase are shown in the red curve of Fig.~7. 
Form factors, as expected, become more important at higher energies,
as indicated by ``no form factors" curve.
The ``no rho" is obtained by taking the limit 
$m_\rho \rar \infty$ in the calculation and tends to that labelled $A_c$ in Fig.~5.
The most prominent feature of the full phase is that it
now has a negative value at threshold, showing that contributions
from light intermediate resonances are important. 
So far, the rho has just been treated as a point-like particle,
however its width, associated with two-pion intermediate states,
is a new source of complex amplitudes and was included in Fig.~7(right) 
 with the label ``fat rho", which compared with the point-like $\r$ show the relevance of including off-shell effects in this vertex.

\section{Final remarks}
We have presented results for the decay $D^+ \rar K^- \p^+ \p^+$
and shown that final state  interactions are visible in data.
It is already clear that
hadronic processes occurring between the primary weak decay 
and asymptotic propagation to the detector do play a key role in 
shaping experimental results.
Although derived from a single instance, the patterns of hadronic interpolation are quite general 
and it is fair to assume that
this conclusion can be extended to other processes. 

\vspace*{-4mm}
\Acknowledgements
PCM would like to thank the organizers for the nice conference and for the opportunity to present a talk. PCM was supported by FAPESP, process 09/50634-0.

\end{document}